\documentclass[aps,pre,showpacs,twocolumn]{revtex4-1}
\usepackage{amsmath}
\usepackage{graphicx}

\usepackage{graphicx}

\begin{document}

\title{Interacting fluids in an arbitrary external field}

\author{Benaoumeur Bakhti}
\affiliation{Department of Physics, University of Mascara, Mascara 29000, Algeria}

\date{\today}

\begin{abstract}
We present new method for studying the equilibrium properties of interacting fluids in an arbitrary external filed. The method is valid in any dimension and it yields an exact results in one dimension. Using this approach, we derive a recurrence relation for the pair distribution function
of a three dimensional in-homogeneous fluids, constitute of spherical molecules 
with arbitrary nearest neighbour interaction that extends to two molecules diameter. By integrating this recurrence relation, we get an explicit expressions for the entropy and free energy functionals as a functionals of the density and the pair distribution function. We show that for one dimensional systems, our results coincide exactly with previously derived one using a completely different approach.

\end{abstract}



\maketitle

In-homogeneous fluids of interacting particles such as sticky or ionic fluids,
exhibit wide and complex phenomena. Though several experimental works have well characterized the complex behaviour of inhomogeneous interacting fluids, a complete microscopic
description of the observed behavior is still missing. Driven by their
widespread applications such as in micro-fluidic devices \cite{Sengupta:2013}
and drag delivery in nanobiotechnology \cite{Fanun:2010},
a great deal of attention has been devoted to interacting fluids during the
last years \cite{Miller/Frenkel:2004a,*Miller/Frenkel:2004b,Buzzaccaro/etal:2007,Hansen-goos/Wettlaufer:2011}. At the aim of understanding their properties at the nanoscale,
different analytical and computational tools have been suggested to investigate both the
equilibrium and the non-equilibrium behavior of interacting fluids.

On the microscopic scale, density functional theory (DFT) \cite{Tarazona:1985,Denton/Ashcroft:1985,Curtin/Ashcroft:1985,Evans:1992,Lowen:1994,Cuesta/Martinez:1997a,Tutschka/Kahl:2000,Tarazona/etal:2008,HansenGoos/Mecke:2009,Lutsko:2010} provides a powerful tool to
investigate the equilibrium properties of interacting fluids. But despite many successes, equilibrium DFT suffers from the fact that good approximations exist only for hard particles (via the fundamental measures approach \cite{Rosenfeld:1989,Rosenfeld/etal:1996,*Rosenfeld/etal:1997,Tarazona:2000,Lafuente/Cuesta:2002a,*Lafuente/Cuesta:2004}) or weakly
interacting particles. The problem of constructing an approximate density functional for more general interactions remains open. 

Due to the fact that much of the rich behavior of real complex fluids stems from the interaction potentials and the spherically anisotropic shapes of constituent particles, we are interested in this work to further extend the DFT for fluids of hard bodies beyond the hard sphere model along the first line, namely introducing an arbitrary nearest neighbour interactions that extends to two molecules size.

Inspired partially from the work of Percus \cite{Percus:1989} on one dimensional hard rods, 
and also from a probabilistic modeling introduced for one dimensional lattice hard rods,
first in \cite{Robledo/Varea:1981} and then extended to interacting rods on a lattice in \cite{Buschle/etal:2000a,
*Buschle/etal:2000b,Bakhti/etal:2012,Bakhti/etal:2013a,Bakhti/etal:2015c}, we derive a recurrence relation for the PDF that is valid in any dimension. Using this recurrence relation, we get also an explicit expressions for entropy and free energy as a functionals of the density and the PDF.
The merit of our approach is that the total correlation function (TCF) or the
radial distribution function (RDF) can be determined directly from the PDF without
further differentiation of the free energy functional (which is in general tedious
because of the complex structures of the free energy functionals). The radial
distribution function can be determined directly from neutron scattering
experiments, and this allows to compare results to the existing one from experiment.
\begin{figure}[b]
\centering
\includegraphics[scale=0.10]{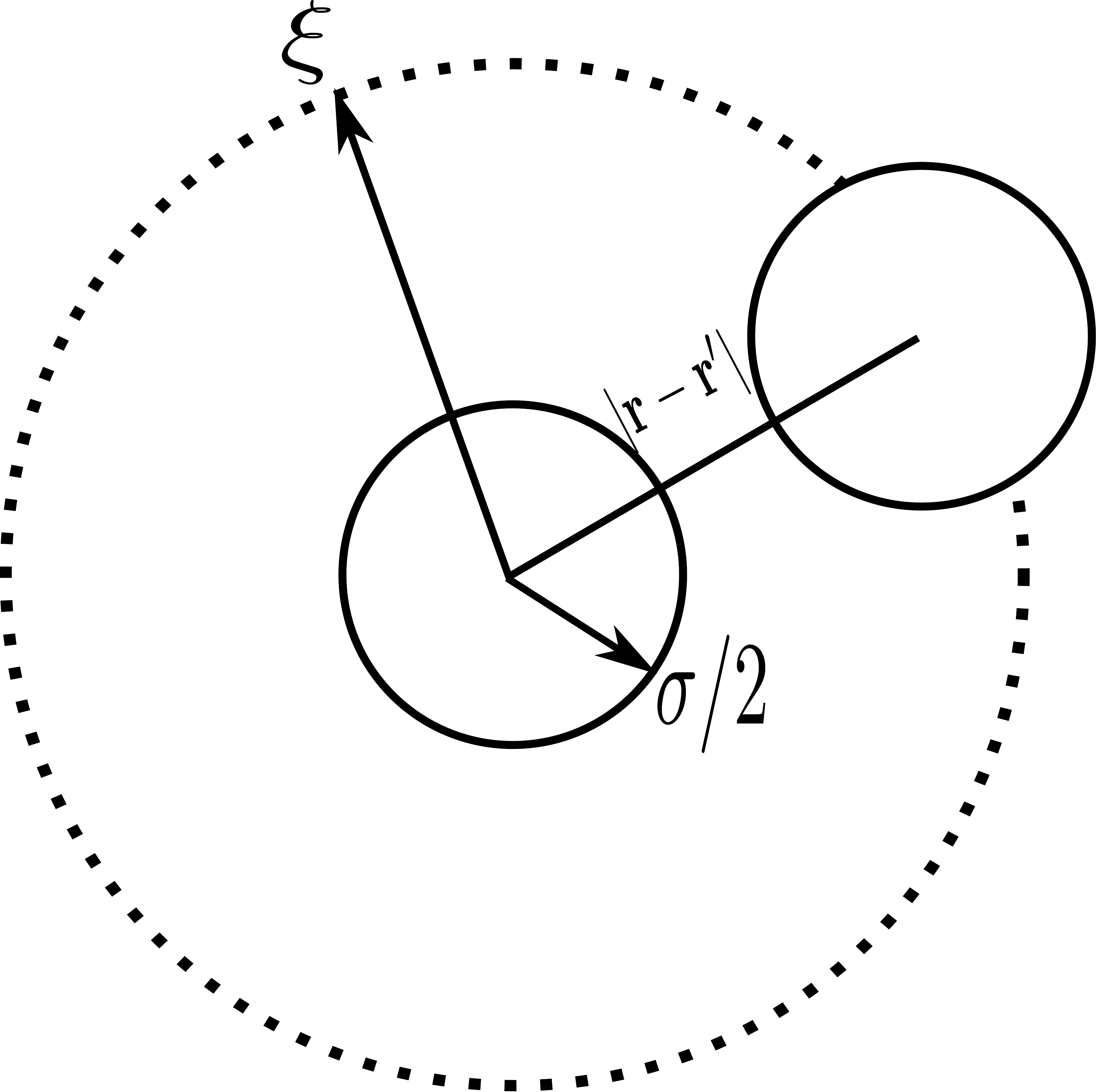}
\caption{\label{fig:colloid} Two hard spheres with diameter $\sigma$ at distance $|\mathbf{r}-\mathbf{r'}|$. If  $|\mathbf{r}-\mathbf{r'}|>\xi$ ($\xi$ is the range of the interaction) the two spheres are not interacting any more. $\xi$ is limited to $\xi <2\sigma$ in order that the interaction does not extend to the next nearest neighbour sphere.}
\end{figure}

The systems we are considering are inhomogeneous fluids of spherically shaped molecules such as colloids,
interacting through a nearest neighbour interaction $\phi$, and they are subject to an arbitrary external field $V_{ex}$ such as gravity. 
These systems are modeled by an interacting hard spheres model (Fig~(\ref{fig:colloid})) with interaction potential given by
\begin{equation}\label{eq:1}
\phi(\mathbf{r}_i,\mathbf{r}_j)=\left\lbrace \begin{array}{ll}
\infty                  &   \mbox{for $    |\mathbf{r}_i-\mathbf{r}_j| < \sigma$} \\
\epsilon(|\mathbf{r}_i-\mathbf{r}_j|)   &   \mbox{for $\sigma $ $\le |\mathbf{r}_i-\mathbf{r}_j| \le \xi$} \\
0          &   \mbox{for $    |\mathbf{r}_i-\mathbf{r}_j| > \xi$}
\end{array}\right. 
\end{equation}
$\epsilon(|\mathbf{r}_i-\mathbf{r}_j|)$ is an arbitrary nearest neighbour interaction (which can be repulsive or attractive), $\sigma$ is the hard sphere diameter and $\xi$ is the total range of the interaction. To exclude next nearest neighbour interaction between hard spheres we take the maximum value of $\xi$ to be $2\sigma$.

A quantities of central interest in DFT are the direct correlation functions (DCF).
Generally, the DCF can be determined from the free energy functional via second differentiation with respect to the one point density. If the free energy functional
is not known, it can be determined from the Ornstein-Zernike (OZ) equation. But here a closer relation is required between the two unknowns (DCF and TCF) of the OZ equation. Different closer relations have been proposed  to solve the OZ equation, include the Percus-Yevick \cite{Percus/Yevick:1958}, Hypernetted chains \cite{Leeuwen/etal:1959,Rosenfeld/Ashcrof:1979}, Born-Green \cite{Born/Green:1946} and the Mean spherical approximations \cite{Lebowitz/Percus:1966}. The resulting DCF’s can be used as a building block to build the free energy functionals.
We fellow here a different root that does not require any closer relation and that gives
a recurrence relation for the PDF (or similarly the RDF or the TCF). To start, let us consider four probability distributions
\begin{align}\label{eq:2}
&P_{11}=\int \!\! d\mathbf{r}_1d\mathbf{r'}_1 \delta(\mathbf{r}_1-\mathbf{r})\delta(\mathbf{r'}_1-\mathbf{r'})\psi(\mathbf{r}_1,\mathbf{r'}_1)\\
&P_{10}=\int \!\! d\mathbf{r}_1d\mathbf{r'}_1\delta(\mathbf{r}_1-\mathbf{r})\omega(|\mathbf{r'}_1-\mathbf{r'}|,\sigma/2,\xi)\psi(\mathbf{r}_1,\mathbf{r'}_1)\nonumber\\
&P_{01}=\!\!\int \!\! d\mathbf{r}_1d\mathbf{r'}_1\omega(|\mathbf{r}_1\!-\!\mathbf{r}|,\xi,\sigma/2)\delta(\mathbf{r'}_1\!-\!\mathbf{r'})\psi(\mathbf{r}_1,\mathbf{r'}_1)\nonumber\\
&P_{00}=\int\!\! d\mathbf{r}_1d\mathbf{r'}_1\omega(|\mathbf{r}_1\!-\!\mathbf{r}|,\xi,\sigma/2)\nonumber\\
&\hspace{20mm}\times\omega(|\mathbf{r'}_1-\mathbf{r'}|,\sigma/2,\xi)\psi(\mathbf{r}_1,\mathbf{r'}_1)\nonumber
\end{align}
where $\delta$ is the delta function and $\omega$ is a weighted function that takes the value $1$ if there is no particle in the interaction region \cite{comm:int_reg} and $0$ else. It has the form
\begin{align}\label{eq:3}
\omega(a,b,c)=1-\theta(a-b)\theta(c -a)
\end{align}
$\theta$ is the Heaviside step function.
The joint probability distribution $\psi$ appearing in Eq.~(\ref{eq:2}) is the
total probability distribution (for $N$ particles). For simplicity of the
notation we wrote $\psi(\mathbf{r}_1,\mathbf{r'}_1)$ for it and we omitted also integration with 
respect to $N-2$ coordinates of the other particles. The previous probability 
distributions have a simple interpretations: $P_{11}$ is the probability to find a particle 
at position $\mathbf{r}$ and another particle at position $\mathbf{r'}$ simultaneously. By definition, it is the PDF. $P_{10}$ ($P_{01}$) is the probability to find a particle (no particle) at position $\mathbf{r}$ and no particle (a particle) at position $\mathbf{r'}$ in the interaction region \cite{comm:int_reg}. $P_{00}$ is the probability to find no particle at position $\mathbf{r}$ and no particle at position $\mathbf{r'}$ in the interaction region.
From the definition of the density and the PDF, we get the following
expressions for the $P_{ij}$
\begin{align}\label{eq:4}
&P_{11}=\rho^{(2)}(\mathbf{r},\mathbf{r'})\\
&P_{10}=\rho(\mathbf{r}) - \int_{V'}\!\! d\mathbf{r'}\rho^{(2)}(\mathbf{r},\mathbf{r'})\nonumber\\
&P_{01}=\rho(\mathbf{r}) - \int_{V}\!\! d\mathbf{r}\rho^{(2)}(\mathbf{r},\mathbf{r'})\nonumber\\
&P_{00}= 1 - \int_{V_T}\!\!d\mathbf{r'}\rho(\mathbf{r'}) + \int_{V,V'}\!\! d\mathbf{r}_1\hspace{0.5mm}d\mathbf{r'}_1\rho^{(2)}(\mathbf{r}_1,\mathbf{r'}_1)\nonumber
\end{align}
$V$ is volume of the hard sphere centered at position $\mathbf{r}$, $V'$ 
is the the volume of the interaction region surrounding the hard sphere (from $\sigma/2$
to $\xi$) and $V_T$ is the
volume of the sphere of radius $\xi$ centered at $\mathbf{r}$. The result of integration depends on the reference point $\mathbf{r}$. For simplicity, we will omit the board of integration from the equations in the following.

The probability distributions in Eq.~(\ref{eq:4}) can be determined also from the
Boltzmann distribution $\psi\!\!=\!\!\exp(-\beta\mathcal{H})/Z$, where $\mathcal{H}$ is the total Hamiltonian, $Z$ is the partition function, $\beta=1/k_BT$ is the inverse temperature and $k_B$ the Boltzmann constant. 
The probability distributions can be written as
\begin{align}\label{eq:5a}
&P_{00}= \frac{1}{Z}\!\!\int \!\!d\mathbf{R} e^{-\beta \mathcal{E}}\\
&P_{10}= \frac{1}{Z}e^{-\beta V_{ex}(\mathbf{r})}\!\!\int \!\!d\mathbf{R} e^{-\beta\mathcal{E}_{\mathbf{r}}}e^{-\beta \mathcal{E}}\nonumber\\
&P_{01}= \frac{1}{Z}e^{-\beta V_{ex}(\mathbf{r'})}\!\!\int \!\!d\mathbf{R}e^{-\beta\mathcal{E}_{\mathbf{r'}}}e^{-\beta \mathcal{E}}\nonumber\\
&P_{11}= \frac{1}{Z}e^{-\beta(\phi(\mathbf{r},\mathbf{r'})+V_{ex}(\mathbf{r})+V_{ex}(\mathbf{r'}))}\!\!\int \!\!d\mathbf{R}e^{-\beta\mathcal{E}_{\mathbf{r}}}e^{-\beta\mathcal{E}_{\mathbf{r'}}}e^{-\beta \mathcal{E}}\nonumber
\end{align} 
where $d\mathbf{R}=d\mathbf{r}_1\ldots d\mathbf{r}_{N-2}$, $\mathcal{E}$ is the energy of the system without the two particles at positions $\mathbf{r}$ and $\mathbf{r'}$ and $\mathcal{E}_{\mathbf{r}}$ ($\mathcal{E}_{\mathbf{r'}}$) is the interaction energy (that includes the external field) between the particle at position $\mathbf{r}$ ($\mathbf{r'}$) and its nearest neighbour molecules. By comparing these Boltzmann distributions, we made the following approximation
\begin{align}\label{eq:5}
P_{00}P_{11}\simeq P_{10}P_{01}e^{-\beta\phi(\mathbf{r},\mathbf{r'})}
\end{align}
that is valid in any dimension and exact in one dimension. Note that for a fixed configuration 
of the system (position of the other $N-2$ particles are fixed), Eqs.~(\ref{eq:5}) is exact in any dimension.

Inserting Eqs.~(\ref{eq:4}) of $P_{ij}$ in (\ref{eq:5}) we get a recurrence relation for the PDF
\begin{align}\label{eq:6}
\rho^{(2)}(\mathbf{r},\mathbf{r'})&=\\
&\frac{\Bigl[\rho(\mathbf{r}) - \int\!\! d\mathbf{r'}\rho^{(2)}(\mathbf{r},\mathbf{r'})\Bigr]\Bigl[\rho(\mathbf{r'}) - \int\!\! d\mathbf{r}\rho^{(2)}(\mathbf{r},\mathbf{r'})\Bigr]}{e^{\beta\phi(\mathbf{r},\mathbf{r'})}\Bigl[1 - \int\!\! d\mathbf{r}_1\rho(\mathbf{r}_1) + \int\!\! d\mathbf{r}_1d\mathbf{r'}_1\rho^{(2)}(\mathbf{r}_1,\mathbf{r'}_1)\Bigr]}\nonumber
\end{align}
The radial distribution function $g$ and the total correlation function $h$ can be calculated straightforwardly using $g(\mathbf{r},\mathbf{r'})=\rho^{(2)}(\mathbf{r},\mathbf{r'})/\rho(\mathbf{r})\rho(\mathbf{r'})$ and $h(\mathbf{r},\mathbf{r'})=g(\mathbf{r},\mathbf{r'})-1$. 

In addition, the recurrence relation (\ref{eq:6}) permits to determine the 
free energy functional. The letter can be written in the form
\begin{align}\label{eq:7}
\Omega &= \int\!\! d\mathbf{r}d\mathbf{r'}\hspace{0.5mm}\rho^{(2)}(\mathbf{r},\mathbf{r'})\phi(\mathbf{r},\mathbf{r'}) + \int\!\! d\mathbf{r}\hspace{0.5mm}\rho(\mathbf{r})V_{ex}(\mathbf{r})\nonumber \\
&- T\!\!\int\!\! d\mathbf{r}\hspace{0.5mm}\tilde{S} - \int\!\! d\mathbf{r}\hspace{0.5mm}\mu\rho(\mathbf{r})
\end{align}
$\mu$ and $\tilde{S}$ are respectively the chemical potential and the entropy density functional. The condition $\delta\Omega/\delta\rho^{(2)}(\mathbf{r},\mathbf{r'})=0$ \cite{Bakhti:2013} yields
\begin{equation}\label{eq:8}
T\tilde{S} = \int\!\!\phi(\mathbf{r},\mathbf{r'})\hspace{0.5mm}d\rho^{(2)}(\mathbf{r},\mathbf{r'})
\end{equation}
We get an expression for the interaction potential $\phi$ as a functional of the PDF by taking the logarithm of the recurrence relation ~(\ref{eq:6}). The integral in ~(\ref{eq:8}) can be calculated exactly, leading to an explicit expression for the entropy as a functional of the density and the PDF,
\begin{align}\label{eq:9}
&S[\rho,\rho^{(2)}]/k_B =\nonumber\\
& -\!\!\int\!\!d\mathbf{r}\Bigl\{\Bigl[\rho(\mathbf{r}) - \!\!\int\!\! d\mathbf{r'}\rho^{(2)}(\mathbf{r},\mathbf{r'})\Bigr]\ln\Bigl[\rho(\mathbf{r}) - \!\!\int\!\! d\mathbf{r'}\rho^{(2)}(\mathbf{r},\mathbf{r'})\Bigr] \nonumber\\
&-\Bigl[\rho(\mathbf{r}) - \int\!\! d\mathbf{r'}\rho^{(2)}(\mathbf{r'},\mathbf{r})\Bigr]\ln\Bigl[\rho(\mathbf{r}) - \int\!\! d\mathbf{r'}\rho^{(2)}(\mathbf{r'},\mathbf{r})\Bigr] \nonumber\\
& - \Bigl[1 - \int_V\!\! d\mathbf{r}_1\rho(\mathbf{r}_1) + \int\!\! d\mathbf{r}_1d\mathbf{r'}_1\rho^{(2)}(\mathbf{r}_1,\mathbf{r'}_1)\Bigr]\nonumber\\
&\times\ln\Bigl[1 - \int_V\!\! d\mathbf{r}_1\rho(\mathbf{r}_1)+ \int\!\! d\mathbf{r}_1d\mathbf{r'}_1\rho^{(2)}(\mathbf{r}_1,\mathbf{r'}_1)\Bigr] \nonumber\\
&- \int\!\!d\mathbf{r'}\rho^{(2)}(\mathbf{r},\mathbf{r'})\ln\rho^{(2)}(\mathbf{r},\mathbf{r'})\Bigr\}+ f(\rho(\mathbf{r}))
\end{align}
We determined the total constant of integration $f(\rho)$ by taking the ideal gas limit $\phi(\mathbf{r},\mathbf{r'})=0$ and we get
\begin{equation}\label{eq:10}
f(\rho(\mathbf{r})) = \int\!\! d\mathbf{r}\rho(\mathbf{r})\ln\rho(\mathbf{r})
\end{equation}
Combining Eqs.~(\ref{eq:6}), (\ref{eq:7}) , (\ref{eq:9}) and (\ref{eq:10}), the free energy functional is given by
\begin{align}\label{eq:12}
&\Omega[\rho,\rho^{(2)}]=\int\! d\mathbf{r}\Bigl\{\rho(\mathbf{r})\ln\Bigl[\rho(\mathbf{r}) - \int\! d\mathbf{r'}\rho^{(2)}(\mathbf{r},\mathbf{r'})\Bigr]
\nonumber\\
&+\rho(\mathbf{r})\ln\Bigl[\rho(\mathbf{r}) -\!\! \int\! d\mathbf{r'}\rho^{(2)}(\mathbf{r'},\mathbf{r})\Bigr] - \rho(\mathbf{r})\ln\rho(\mathbf{r})\nonumber\\
&-\!\!\Bigl[1 \!-\!\! \int_V\!\! d\mathbf{r}_1\rho(\mathbf{r}_1)\Bigr]\!\!\ln\Bigl[1\! -\!\! \int_V\!\! d\mathbf{r}_1\rho(\mathbf{r}_1)\! +\!\! \int\!\! d\mathbf{r}_1d\mathbf{r'}_1\rho^{(2)}(\mathbf{r}_1,\mathbf{r'}_1)\Bigr] \nonumber\\
& + \left(V_{ex}(\mathbf{r})-\mu\right)\rho(\mathbf{r})\Bigr\}
\end{align}
Other thermodynamic functions can be inferred from the free energy functional~(\ref{eq:12}). The formalism presented here is valid in any dimension. An exact and explicit results can be found only for some specific systems in one dimension such as the sticky hard spheres confined into a one dimensional channel.
In higher dimensions, numerical work is required. The numerical treatment of an in-homogeneous three dimensional colloidal systems with some specific interactions (such as the sticky and square well potentials) is postponed for future work.

For hard spheres with arbitrary finite range interaction confined into a one dimensional channel,
Eq.~(\ref{eq:6}) is reduced to
\begin{align}\label{eq:13}
\rho^{(2)}(y,y')&=\\
&\frac{\Bigl[\rho(y) - \int\!\! dy'\rho^{(2)}(y,y')\Bigr]\Bigl[\rho(y') - \int\!\! dy\rho^{(2)}(y,y')\Bigr]}{e^{\beta\phi(y,y')}\Bigl[1 - \int\!\! dy_1\rho(y_1) + \int\!\! dy_1dy'_1\rho^{(2)}(y_1,y'_1)\Bigr]}\nonumber
\end{align}
The range of variation of $y'$ is $[y+|\sigma|/2,y+|\xi|]$ or differently $[y-\xi,y-\sigma/2]$and $[y+\sigma/2,y+\xi]$. The entropy has the exact form
\begin{align}\label{eq:14}
&S[\rho,\rho^{(2)}]=\nonumber\\
&-\!\!\int\!\! dy\!\left\lbrace\!\!\Bigl[\rho(y) \!-\!\! \int\!\! dy'\rho^{(2)}(y,y')\Bigr]\!\ln\Bigl[\rho(y) \!-\!\! \int\!\! dy'\rho^{(2)}(y,y')\Bigr]\right.\nonumber\\
&+\Bigl[\rho(y) -\!\! \int\!\! dy'\rho^{(2)}(y',y)\Bigr]\ln\Bigl[\rho(y) -\!\! \int\!\! dy'\rho^{(2)}(y',y)\Bigr] \nonumber\\
& -\rho(y)\ln\rho(y) +\int\!\! dy'\rho^{(2)}(y,y')\ln\!\rho^{(2)}(y,y')\nonumber\\
&\left.+\Bigl[1 -\!\! \int\!\! dy_1\rho(y_1) +\!\! \int\!\! dy_1dy'_1\rho^{(2)}(y_1,y'_1)\Bigr]\right.\nonumber\\
&\left.\times\ln\Bigl[1 -\!\! \int\!\! dy_1\rho(y_1) +\!\! \int\!\! dy_1dy'_1\rho^{(2)}(y_1,y'_1)\!\Bigr] \right\rbrace
\end{align}
which coincides exactly with Percus result \cite{Percus:1989}. In addition, we get an exact expression free energy functional
\begin{align}\label{eq:15}
&\Omega[\rho,\rho^{(2)}]=\int\!\! dy\Bigl\{\rho(y)\ln\Bigl[\rho(y) -\!\! \int\!\! dy'\rho^{(2)}(y,y')\Bigr]
\nonumber\\
&+\rho(y)\ln\Bigl[\rho(y) -\!\! \int\!\! dy'\rho^{(2)}(y',y)\Bigr]- \rho(y)\ln\rho(y)\nonumber\\
&-\!\Bigl[1 \!-\!\! \int\!\! dy_1\rho(y_1)\Bigr]\!\!\ln\Bigl[1 -\!\! \int\!\! dy_1\rho(y_1)\! +\!\! \int\!\! dy_1dy'_1\rho^{(2)}(y_1,y'_1)\Bigr] \nonumber\\
& + \left(V_{ex}(y)-\mu\right)\rho(y)\Bigr\}
\end{align}
The recurrence relation ~(\ref{eq:13}) can be solved exactly in the case of sticky core interactions
where the hard spheres interact through a strong attractive interaction if they are in contact ($\xi=\sigma$). Eq. ~(\ref{eq:13}) is reduced to
\begin{align}\label{eq:16}
\rho^{(2)}(y,y')&=\\
&\frac{\Bigl[\rho(y) - \rho^{(2)}(y,y')\Bigr]\Bigl[\rho(y') - \rho^{(2)}(y,y')\Bigr]}{e^{\beta\phi(y,y')}\Bigl[1 - \int_{y-\sigma/2}^{y+\sigma/2}\! dx\rho(x) + \rho^{(2)}(y,y')\Bigr]}\nonumber
\end{align}
giving rise to a quadratic equation in $\rho^{2}$ with an exact and explicit solution
\begin{align}\label{eq:17}
\rho^{(2)}(y,y')&=\\
&\frac{1}{2\eta}\Bigl[K- \Bigl[K^{2} - 4\eta(\eta + 1)\rho(y)\rho(y')\Bigr]^{1/2}\Bigr]\nonumber
\end{align}
with
\begin{equation}\label{eq:18}
K=1+e^{-\beta\phi(y,y')}\Bigl(\rho(y)+\rho(y')\Bigr)- \int_{y-\sigma/2}^{y+\sigma/2}\!\! dx\rho(x)
\end{equation}
and $\eta = e^{-\beta\phi(y,y')}-1$. $y$ and $y'$ are related by $y'-y=\sigma$.

From ~(\ref{eq:16}), we infer an exact and explicit expression for the entropy functional
\begin{align}\label{eq:19}
&S/k_B =\int\!\! dy\Bigl\{-\Bigl[\rho(y) - \rho^{(2)}(y,y')\Bigr]\ln\Bigl[\rho(y) - \rho^{(2)}(y,y')\Bigr] \nonumber\\
&-\Bigl[\rho(y') - \rho^{(2)}(y,y')\Bigr]\ln\Bigl[\rho(y') - \rho^{(2)}(y,y')\Bigr] \nonumber\\
&- \!\!\Bigl[1 \!-\!\! \int\!\! dx\rho(x)\! + \!\rho^{(2)}(y,y')\Bigr]\!\ln\Bigl[1\! -\!\! \int\!\! dx\rho(x) \!+ \!\rho^{(2)}(y,y')\Bigr]\nonumber\\
 &- \rho^{(2)}(y,y')\ln\Bigl[\rho^{(2)}(y,y')\Bigr] +\rho(y)\ln\rho(y)\Bigr\}
\end{align} 
and for the free energy functional
\begin{align}\label{eq:20}
\Omega[\rho,\rho^{(2)}]&=\int\!\! dy\Bigl\{\rho(y)\ln\Bigl[\rho(y) - \rho^{(2)}(y,y')\Bigr]
\nonumber\\
&+\rho(y')\ln\Bigl[\rho(y') - \rho^{(2)}(y,y')\Bigr]\nonumber\\
&-\Bigl[1 -\!\! \int\!\! dx\rho(x)\Bigr]\ln\Bigl[1 -\!\! \int\!\! dx\rho(x) + \rho^{(2)}(y,y')\Bigr] \nonumber\\
&- \rho(y)\ln\rho(y) + \left(V_{ex}(y)-\mu\right)\rho(y)\Bigr\}
\end{align}

For three dimensional sticky core fluids, the volume integral of the PDF of the rhs of eq~ (\ref{eq:6}) is reduced to a surface integral over the surface of the hard sphere located at position $\mathbf{r}$
\begin{align}\label{eq:21}
\rho^{(2)}&(\mathbf{r},\mathbf{r'})=\\
&\frac{\Bigl[\rho(\mathbf{r}) - \int_{S}\!\! d\mathbf{r'}\rho^{(2)}(\mathbf{r},\mathbf{r'})\Bigr]\Bigl[\rho(\mathbf{r'}) - \int_{S}\!\! d\mathbf{r}\rho^{(2)}(\mathbf{r},\mathbf{r'})\Bigr]}{e^{\beta\phi(\mathbf{r},\mathbf{r'})}\Bigl[1 - \int_{V}\!\! d\mathbf{r}_1\rho(\mathbf{r}_1) + \int_{S}\!\! d\mathbf{r'}\rho^{(2)}(\mathbf{r},\mathbf{r'})\Bigr]}\nonumber
\end{align}
with $|\mathbf{r'}-\mathbf{r}|=\sigma$. $S$ and $V$ are respectively the surface and volume of the hard sphere centered at position $\mathbf{r}$. Entropy and free energy functionals are given exactly by 
the same expressions as in ~(\ref{eq:9}), ~(\ref{eq:10}) and ~(\ref{eq:12}) respectively with the volume integral of the PDF is replaced by a surface integral as in Eq.~(\ref{eq:21}).

In conclusion, we presented here a new approach for studying the thermodynamics of an interacting fluids in an arbitrary external filed. Based on the new approach, we got a recurrence relation for the 
PDF of inhomogeneous fluids with general finite range nearest neighbour interactions. 
We shown that an explicit expressions for entropy and free energy functional as a functionals of the density and the PDF can be derived from this recurrence relation. 
For spherical molecules confined into a one dimensional channel, our results coincide exactly
with the exact results of Percus \cite{Percus:1989}. This approach yields an exact results in one dimension because in this case Eq.~\ref{eq:5} is exact. In the limit of sticky core particles confined into a one dimensional channel,
the recurrence relation admits an exact and explicit solution. Extension of this work to mixture of interacting fluids in an inhomogeneous field is in progress. In the letter case, the molecules
are distinguished by their sizes and their species. Further possible extension of this work that will be considered in the future, is the interacting molecules that have orientational degrees of freedom, such as hard rods, hard platlets and hard ellipsoids.\\

We thank M. Kr\"uger, A. Kl\"umper and M. Karbach for very valuable
discussions.

\bibliographystyle{apsrev4-1}
\bibliography{dft}

\begin{thebibliography}{36}
\bibitem [{com()}]{comm:int_reg}%
  \BibitemOpen
  \href@noop {} {}\bibinfo {note} {The interaction region is the volume
  serounding the molecule (hard sphere) and that extends from $\sigma/2$ to
  $\xi$}\BibitemShut {NoStop}%
\end{thebibliography}

\end{document}